\documentclass{elsart}
\usepackage{amssymb}

\begin{document}

\begin{frontmatter}

\title{High temperature periodic BEC in deuterated metals}

\author{J.S.Brown}
\ead{j.brown3@physics.ox.ac.uk}
\address{University of Oxford, Clarendon Laboratory, Parks Rd., OX1 3PU, England}

\begin{abstract}
The standard Fr\"ohlich model of the interaction of band states with phonons is applied to the case of deuterons solved in metals. A simple variational solution of this model predicts at least one highly coherent state of the deuteron-phonon system having a total energy significantly lower than that of uncorrelated interstitial deuterons and only thermal phonons, and that this phase may actually be exhibited at room temperature and high levels of deuteration.
\end{abstract}

\end{frontmatter}

\section{Introduction}
A recent attempt by K\'alm\'an and Keszthelyi \cite{Kal05} to explain the anomalously large fusion cross-section observed when low energy deuterons traverse Pd foils at low energy was found insufficient to account for the full magnitude of the reported effect. The present work was motivated by the observation that their model is based upon an {\it a priori} transformation of the phonon-deuteron vertex of the Fr\"ohlich Hamiltonian into an effective deuteron-deuteron interaction, and is therefore unable to take account of any collective effects that might arise from coherent phonon excitations.
\section{Model}
The essential physics of hydrogen nucleii solved in metal lattices can be captured by a single-band Fr\"ohlich model because the lowest states available form a band, albeit a very narrow one \cite{Pus83}, that is separated from the excited states by a constant band gap that is well in excess of both $kT$ and $\omega_D$.
\\
The Hamiltonian for a variable number of deuterons solved on a lattice in exchange equilibrium with the gas phase is
\begin{equation}
\label{hf}
H = \epsilon_k c_k{^\dag}c_k +  M_q (b_q + b^{\dag}_{-q}) c^{\dag}_{k+q} c_k  + U_q c^{\dag}_{k_1+q}  c^{\dag}_{k_2-q} c_{k_2} c_{k_1} + \omega_q b^{\dag}_q b_q
\end{equation}
$\epsilon_k$ is the band energy of a deuteron of crystal momentum $k$ and is negligible in comparison with the other terms, $c_k^{\dag}$ is the raising operator for a deuteron of wavevector $k$, $b_q^{\dag}$ is the raising operator for a phonon of wavevector $q$. We have neglected spin and band indices for the sake of clarity.
\\
According to the Thomas-Fermi model \cite{Cha83} of electronic screening in metals
\begin{equation}
U_q \approx { e^2 \over q^2 + 4\pi e^2 N(\epsilon_F)}
\end{equation}
The bosonic character of the deuterons is preserved in the metallic environment because the accompanying electronic cloud is made of very many small contributions with zero average spin.
\\
Consistency with the classical model of the dielectric response of the lattice, and consideration of the fact that the retarded interactions will be subject to exactly the same electronic screening response as the instantaneous Coulomb interaction implies that
\begin{equation}
\label{mq}
|M_q|^2 \approx \omega_q U_q
\end{equation}
The phonon vertex interaction is clearly strongest when the deuterons equally populate two distinct states separated by the momentum corresponding to half a reciprocal lattice vector, $\vec G/2$ .
\\
 We denote the state having $N_{dd}$ spin-up deuterons with $\vec k=0$, an equal number of spin-down deuterons with  $\vec k = \vec G/2$ and $N_p$ phonons with $\vec q = \vec G/2$ by 
$|N_{dd},0,0,0,0, N_{dd}, N_p\rangle$. The subspace spanned by this state together with the phonon-coupled state $|N_{dd}-1,0,0,0,1,N_{dd}, N_p-1\rangle $ has the explicit Hamiltonian:
\begin{equation}
\label{h2}
\mathbf{H} = \left ( \begin{array}{cc} 2N_{dd}^2 U_0 + N_p \omega_D & \sqrt{N_p} N_{dd} M_{G/2} \\ \sqrt{N_p} N_{dd} M_{G/2}  & (2N_{dd}^2-2N_{dd}+2) U_0 + (N_p -1)\omega_D \end{array} \right )
\end{equation}
where $\omega_D$ is the Debye energy of the acoustic phonon branch.
\section{Numerical estimate}
With the value of $N(\epsilon_F) = 0.47$ states/(eV Pd) for PdH reported in \cite{Cha83} we have $U_0 \approx 0.8$eV and $U_{G/2} \approx 0.4$eV. For the acoustic branch in Pd, $\omega_D \approx 0.03 $eV.
Substituting the dimensionless parameters i) mean deuterons-per-site $\lambda = {2N_{dd} / N_c}$ and ii) mean squared phonons-per-site $\nu^2 = {N_p / N_c}$ - and using (2) we obtain in the limit of large $N_c$, in eV units:
\begin{equation}
\mathbf{H} = N_c \left ( \begin{array}{cc} 0.4\lambda^2 + 0.03 \nu^2 & 0.18\nu \lambda \\ 0.18\nu \lambda  & 0.4\lambda^2 + 0.03\nu^2 \end{array} \right )
\end{equation}
In view of the variational principle, this calculation represents an upper bound for the lowest energy $\epsilon_-$. If the subspace is expanded to include all even values of $N_p$, for example, the splitting caused by the phonon-interaction term is found to be about twice as large as that given by (\ref{h2}) above and the corresponding site energies (in eV) are:
\begin{equation}
\label{epm}
\epsilon_{\pm} = 0.4\lambda^2 + 0.03\nu^2 \pm 0.36\nu \lambda
\end{equation}
With a chemically achievable $\lambda = 1.0$, the lowest value of $\epsilon_{-} \approx -2$eV is achieved when $\nu \approx 6$. This represents a significant ($\gg kT$) gain in energy relative to (normal) uncorrelated interstitial deuteron states without coherent phonon excitation. It should also be noted that if the metal is in exchange equilibrium with the $D_2$ gas phase, (\ref{epm}) predicts a spontaneous take-up of additional deuterons once a certain threshold value of $\lambda$ has been reached. 
\\

\section{Conclusion}
We conclude that coherent phonon-deuteron interactions  may dominate the Coulomb repulsion energy in certain deuterated metal regimes, even allowing for the phonon energy invested in the highly-excited phonon mode. We also believe that the principle outlined here, which is closely related to the well-known Peierls instability, may be applicable to other situations involving high concentrations of light charged bosonic interstitials in rigid lattices.

\end{document}